\documentclass[twocolumn,showpacs,amsmath,amstex,amssymb,mathfonts,aps,prl]{revtex4}
\usepackage{graphicx,epstopdf,verbatim}
\usepackage{float, color}

\begin{document}
\title{Numerical Study of a Many-Body Localized System Coupled to a Bath}
\author{Sonika Johri$^{1}$, Rahul Nandkishore$^{2}$,  R. N. Bhatt$^{1, 3}$}
\affiliation{$^1$ Department of Electrical Engineering, Princeton University, Princeton, NJ 08544}
\affiliation{$^2$ Princeton Center for Theoretical Science, Princeton University, Princeton, NJ 08544}
\affiliation{$^3$ School of Natural Sciences, Institute for Advanced Study, Princeton, NJ 08540}

\begin{abstract}
We use exact diagonalization to study the breakdown of many-body localization in a strongly disordered and interacting system coupled to a thermalizing environment. We show that the many-body level statistics cross over from Poisson to GOE, and the localized eigenstates thermalize, with the crossover coupling decreasing with the size of the bath in a manner consistent with the hypothesis that an infinitesimally small coupling to a thermodynamic bath should destroy localization of the eigenstates. However, signatures of incomplete localization survive in spectral functions of local operators even when the coupling to the environment is sufficient to thermalize the eigenstates. These include a discrete spectrum and a gap at zero frequency. Both features are washed out by line broadening as one increases the coupling to the bath. We also determine how the line broadening scales with coupling to the bath. 
\end{abstract}
\pacs{78.40.Pg, 71.23.An, 71.30.+h, 72.80.Ng}
\date{\today}
\maketitle


%
Isolated quantum systems with quenched disorder can enter a `localized' regime where they fail to ever reach thermodynamic equilibrium \cite{anderson}. While we have an essentially complete understanding of localization in non-interacting systems \cite{anderson}, the theory of many-body localization (MBL) is still under construction \cite{agkl, Mirlin, BAA, oganesyan, Prosen, pal, Imbrie, LPQO, Bauer, Pekker, Vosk, narrowbath, Bahri, Lbits, serbyn, Abanin, QHMBL, Sid, Bauer2, arcmp, altmanreview}. Numerical investigations using exact diagonalization \cite{oganesyan, pal, serbyn} {\it do} indicate that all eigenstates of a strongly interacting disordered system can be localized. Most of the theoretical research so far has been in the limit of a perfectly isolated system. However, experimental tests of MBL (\cite{shahar, demarco}) will always include some finite coupling to the environment. What then can we expect to see in experiments designed to probe many body localization? 

A recent theory of MBL systems weakly coupled to heat baths proposed that while eigenstates are delocalized by an infinitesimally weak coupling to a heat bath, signatures of localization persist in spectral functions of local operators for weak coupling to a bath \cite{rahul}. This theory has yet to face stringent numerical tests. Moreover, it did not discuss the spectral functions of the physical degrees of freedom, the quantities of direct relevance for experiments, focusing instead on the spectral functions of certain localized integrals of motion that are believed to exist \cite{serbyn, Imbrie, Lbits}, but which are related to the physical degrees of freedom by an unknown unitary transformation. This work directly addresses these issues. 

\begin{figure*}
\hspace*{-5mm}
\includegraphics[width=14cm, height=7cm]{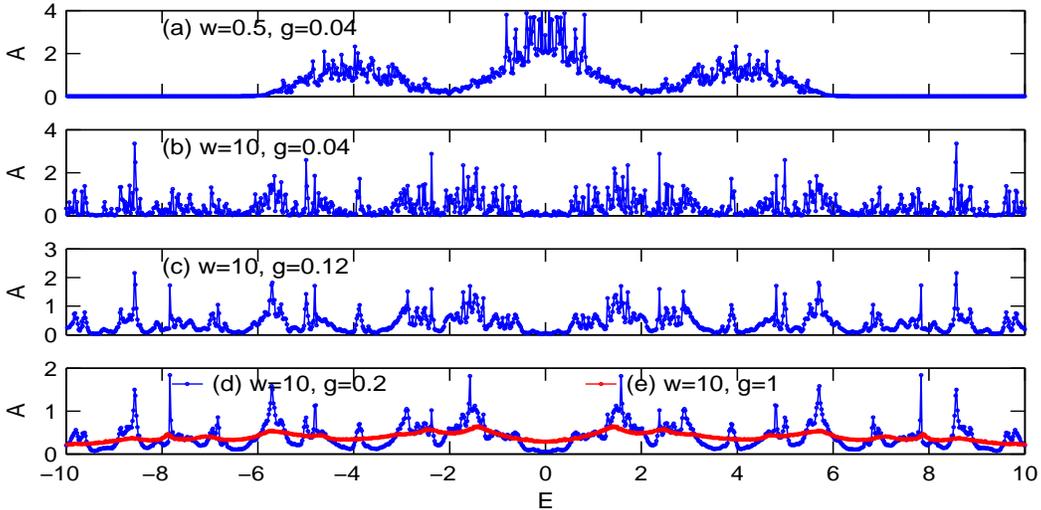}
\vspace{-0.2cm}
\caption[]{(Color online) Evolution of the spectrum of a ($p$-bit) spin flip operator  as coupling $g$ to the bath increases in a typical disorder realization with disorder strength $w$. Results are for a system with periodic boundary conditions containing $N=7$ spins coupled to a bath with $N_b=7$ spins, and are averaged over all spins in the system. The top figure shows the spectral function in the low disorder delocalized phase, whereas the other figures show the spectral function of a system that would be localized if perfectly isolated, but which is coupled to a bath with a coupling $g$. The inhomogeneity of the spectral function and the existence of a hierarchy of gaps (especially the zero frequency gap) are diagnostics of localization. As $g$ is increased (b-d), the structure in the spectral function is gradually washed out, giving a crossover to thermalization. Note that the eigenstates become effectively thermal for $g > 0.09$, according to the finite size scaling (Fig. 3, 4), but the local spectral functions retain signatures of localization until $g$ becomes comparable to the characteristic energy scales in the system ($g \approx 1$). }
\label{fig:pbits}
\vspace{-10pt}
\end{figure*}
We use exact numerical diagonalization to establish the behavior of many body localized systems weakly coupled to heat baths. 
We show that coupling $g$ to a bath results in a crossover from Poisson to Gaussian orthogonal ensemble (GOE) eigenvalue statistics, which becomes exponentially steeper with increasing bath size. A similar rapid crossover to thermalization is seen in the eigenstates. However, the prospect for seeing MBL in experiments is still realistic because signatures of incomplete localization remain in the spectral functions of local (in real space) operators. 
Indeed, we find that the spectral functions of the microscopic degrees of freedom look completely different in the localized and thermal phases (see Fig. 1). The thermal phase has a continuous spectrum whereas the local spectral function in the localized phase is discrete, with a hierarchy of gaps, and a gap at zero frequency that survives even after spatial averaging. Increasing $g$ causes lines to broaden and fill in these gaps. However, as long as the typical line broadening is less than the largest gaps, gap-like features remain. 
 Our work also reveals how the line broadening scales with $g$.  


{\it The model}: We choose the antiferromagnetic Heisenberg spin-$1/2$ chain with random fields along $z$:
\begin{equation}\label{eq:H_pbit}
H_0=\Sigma_{i=1}^{N-1} 2J \vec{\sigma}_i. \vec{\sigma}_{i+1} + \Sigma_{i=1}^N h_i\sigma_i^z 
\end{equation}
We set the interaction $J=1$. The on-site fields $h_i$ are independent random variables, uniformly distributed between $-w$ and $w$; $w$ measures the disorder strength in the system. This model with periodic boundary conditions has been shown to have a many-body localization transition at $w=7$ in the infinite temperature limit \cite{pal}.

The Hamiltonian in Eq. \ref{eq:H_pbit} is written in terms of the physical degrees of freedom $\sigma$ (`$p$-bits,' in the language of \cite{Lbits}, where $p$=physical). In general, its eigenstates are quite complicated and non-trivial. As shown \cite{Lbits, serbyn}, one can perform a unitary transformation to rewrite $H_0$ in terms of localized constants of motion $\tau_{i}^z$. The $\tau^z_i$ are dressed versions of the $\sigma$ operators, which are localized in real space, with exponential tails, and are referred to in \cite{Lbits} as `$l$-bits' ($l$=localized). A unitary transformation to this `$l$-bit' basis can always be performed, if the system is in the regime where all the many body eigenstates are localized. In this $l$-bit basis, the Hamiltonian becomes
\begin{equation}
H_0  =  \Sigma_i \tilde h_i\tau^z_i + \Sigma_{i,j}\tilde J_{ij}\tau^z_i \tau^z_j
 + \Sigma_n \Sigma_{i,j,\{k\}}K^{(n)}_{i\{k\} j}\tau^z_i\tau^z_{k_1}...\tau^z_{k_n} \tau^z_j ~.
 \label{eq:H_u}
\end{equation}

The values of the coefficients  $\tilde h, \tilde J$ and $K^{(n)}_{\{k\}}$ will depend upon the parent Hamiltonian (1), although these coefficients all fall off exponentially with distance. The eigenstates of (2) are just products of $\tau_{i}^{z}$.

Motivated by the representation (2) of the Hamiltonian (1), it is instructive to consider the simpler Hamiltonian
\begin{equation}
H^{(l)}_{0}=\Sigma_{i=1}^{N-1} 2\tilde J_i \tau_i^{z} \tau_{i+1}^{z} + \Sigma_{i=1}^N \tilde h_i^{}\tau_i^{z} 
\end{equation}
where the $\tilde h_i$ and $\tilde J_i$ as independent random variables taken from a log-normal distribution with $\langle \ln \tilde h \rangle = 0$ and $\langle \ln^2 \tilde h \rangle = w^2$, and similarly for $\tilde J$. We take $w=0.5$ and work with open-boundary conditions. This Hamiltonian also has the feature that eigenstates are product states of $\tau^z$, and is simpler to work with numerically. 

For the bath, we use a non-integrable Hamiltonian that has been recently studied \cite{kim}. It consists of $N_b$ interacting spins with the Hamiltonian:
\begin{equation}
H_{bath}=\Sigma_{i=1}^{N_b-1} 2J_b S_i^z S_{i+1}^z + \Sigma_{i=1}^{N_b} h_b S_i^z + \Sigma_{i=1}^{N_b} g_b S_i^x 
\end{equation}
While using open boundary conditions, we add a boundary term $J_b (S_1^z + S_{N_b}^z)$ to $H_{bath}$. We use $J_b=1$, $h_b=0.8090$ and $g_b=0.9045$, values for which $H_{bath}$ has been numerically shown by \cite{kim} to have fast entanglement spreading. (We use periodic boundary conditions only for $p$-bits with $N_b=N$.)

The interaction between the system and bath should be local for both $p$- and $l$-bits. We first study $l$-bit eigenstates, choosing the coupling:
\begin{equation}
H_{int}=g \Sigma_{i=1}^{N-1}  \tau_i^+ \tau_{i+1}^- S_{i+(N_b-N)/2}^x + h.c.
\end{equation}
Later we examine $p$-bit spectra, using the coupling 
\begin{equation}
H_{int}=g \Sigma_{i=1}^{N-1}  \sigma_i^+ \sigma_{i+1}^- S_{i+(N_b-N)/2}^x + h.c.
\end{equation}
The total Hamiltonian is thus $H^{(l)/ (p)}_{T}=H_0+H_{bath}+H_{int}$, where $H_0$ and $H_{int}$ are given by Eq. (3) and (5) in the first part of this work, and by Eq. (1) and (6) in the latter part of this work. We will indicate the transition clearly in the text. We use open boundary conditions except where periodic boundaries are explicitly mentioned.

\begin{figure}[htb]
\includegraphics[width=0.8\linewidth]{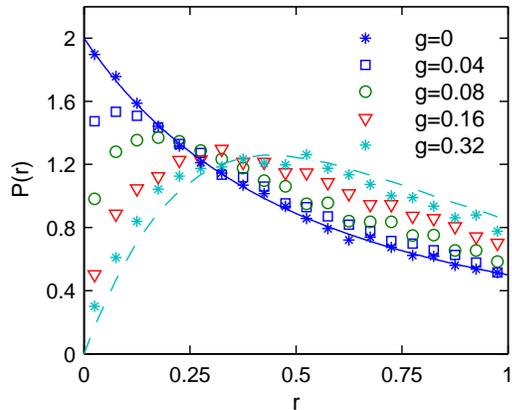}
\vspace{-5mm}
\caption[]{(Color online) Crossover from Poisson to Gaussian statistics in the $l$-bit Hamiltonian as $g$ is increased. Results are for a system with $N=8$ spins and bath with $N_b=7$ spins averaged over $\sim 50,000$ eigenstates obtained from several disorder configurations. The dark blue solid line is the Poisson distribution expected for localized systems, and the light blue dashed line is the GOE distribution expected for thermalizing systems.}
\label{fig:level_space}
\vspace{-15pt}
\end{figure}

We start by analyzing the breakdown of localization when the $l$-bit Hamiltonian (3) is coupled to a bath according to (5), by examining the many-body eigenvalue statistics as $g$ is increased from $0$. We perform exact diagonalization on a system with $N=8$ spins coupled to $N_b=7$ spins in the bath.  The many body level-spacing is $\Delta_n=|E_n-E_{n-1}|$, where $E_n$ is the energy of the $n$th eigenstate.  Following \cite{pal}, we define the ratio of adjacent gaps as $r_n=\min(\Delta_n,\Delta_{n+1})/\max(\Delta_n,\Delta_{n+1})$. We average this over eigenstates and several different realizations of the disorder to get a probability distribution $P(r)$ at a particular value of $g$. In Fig. \ref{fig:level_space}, we show how $P(r)$ evolves from Poisson to GOE like as $g$ is increased. In a localized system we expect that $P(r\rightarrow 0) = 2$, and for a thermalizing system, we expect that $P(r\rightarrow 0)=0$.
The transition from Poisson to GOE statistics happens gradually for this finite size system.  A simple analytical estimate of the characteristic value of $g$ at the crossover point proceeds as follows (see also \cite{rahul}): If $t$ is the bandwidth of the bath and $\delta$ is the many body level spacing in the bath, then the system couples to $\sim t/\delta$ states, with a typical matrix element to each state of order $g \sqrt{\delta/t}$. The coupling to the bath will be effective in thermalizing the system when this matrix element becomes of order the level spacing in the bath, i.e. when $g \sqrt{\delta/t} \sim \delta$. This indicates that the crossover coupling $g_c \sim \sqrt{t \delta}$. 
Since $\delta\sim 2^{-N_b}$, the critical value of $g$ is expected to scale as $g_c \sim \exp(-N_b\log(2)/2)\sim \exp(-0.345N_b)$. 

\begin{figure}[htb]
\includegraphics[width=\linewidth]{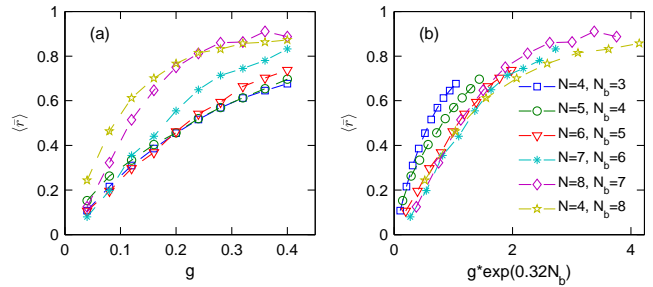}
\vspace{-6mm}
\caption[]{(Color online) (a) The average of the ratio of adjacent energy gaps $\langle\bar{r}\rangle$ (defined in the text) in the $l$-bit Hamiltonian as $g$ is increased for system sizes $N=4,5,6, 7, 8$ and $N_b=N-1$.  Data is averaged over $\sim 50,000$ eigenstates obtained from several disorder configurations. (b) Collapse of data in (a) is in good agreement with analytic arguments for the finite size scaling presented in the main text, and depends only on $N_b$.}
\label{fig:r_trans}
\vspace{-10pt}
\end{figure}

\begin{figure}[htb]
\includegraphics[width=\linewidth]{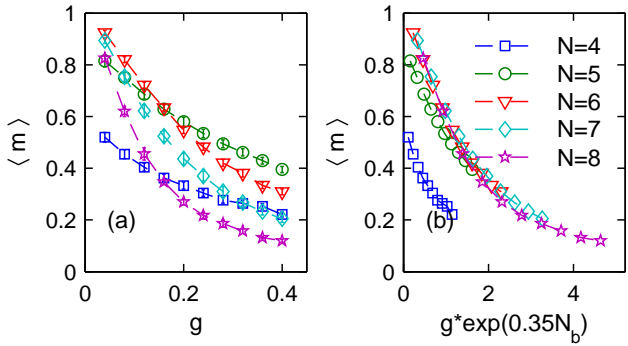}
\vspace{-5mm}
\caption[]{(Color online) (a) Increasing thermalization of the states in the center of the band of the $l$-bit Hamiltonian as $g$ is increased for system sizes $N=4$, $5$, $6$, $7$, $8$ and $N_b=N-1$. $\langle m\rangle$ as defined in the text is measured at the site of the central spin. Data is averaged over $\sim 50,000$ eigenstates obtained from several disorder configurations. (b) Collapse of data in (a) agrees with analytical estimates of finite size scaling for $N_b = N-1 \ge 5$. For a finite size system with $N_b$ spins in the bath, the eigenstates become effectively thermal for $g > \exp(-0.35 N_b)$, implying that eigenstates in the thermodynamic limit become thermal for infinitesimal $g$. }
\label{fig:eth}
\vspace{-5pt}
\end{figure}

To quantitatively compare this crossover estimate to the data, we define $\langle r\rangle=\int P(r)r dr$. After averaging over disorder distributions, $\langle r \rangle$ should be $~0.53$ in the GOE regime and $0.39$ in the localized regime \cite{pal}. It is convenient to define the normalized quantity $\langle \bar{r}\rangle$=($\langle r\rangle-0.39)/0.14$, such that $\langle\bar{r}\rangle = 1$ if the level statistics are GOE and $\langle\bar{r}\rangle = 0$ if they are Poisson. Fig. 3(a) shows how $\langle\bar{r}\rangle$ varies with $g$ for systems of size $N=N_b+1=4,5,6,7,8$. Fig. \ref{fig:r_trans}(b) shows that scaling of the form $g*\exp(aN_b)$ is successful in making the data for different $N_b$ in Fig. \ref{fig:r_trans}(a) collapse onto one curve. Data collapse occurs also for $N=4$ and $N_b=8$, indicating clearly that it is $N_b$ which controls the finite size scaling. We get the best collapse when the constant in the exponential is $\sim 0.32$ which is in good agreement with the analytical estimate $\log (2)/2 \approx 0.345$. This implies that the crossover to thermalization is at a coupling $g_c$ that is exponentially small in system size, so that level statistics become GOE at infinitesimal $g$ in the thermodynamic limit. 
Another test of thermalization is checking whether the eigenstates obey the eigenstate thermalization hypothesis (ETH) \cite{srednicki, Deutsch, Rigol}.  The ETH states that the expectation value of a local operator should be the same in every eigenstate within a small energy window. For a localized system this will not be the case. In Fig. \ref{fig:eth}, we show how eigenstate thermalization sets in as $g$ is increased. We choose an energy window around the center of the band and calculate the standard deviation of the expectation value of $\tau_{N/2}^z$ for all eigenstates within the window. Explicitly, we define 
\begin{equation}
\langle m\rangle= \sqrt{\overline{|<\Psi_i|\tau_{N/2}^z|\Psi_i>|^2 } -\left|\overline{<\Psi_i|\tau_{N/2}^z|\Psi_i>} \right|^2} ; 
\end{equation}
where the overline denotes averaging over an energy window of width $\delta E$ in the middle of the band and $\Psi_i$ is an eigenstate of the coupled system and bath. We choose $\delta E = 0.1$. After averaging over disorder distributions, we expect to find $\langle m\rangle=0$ for a thermalized system. Fig. \ref{fig:eth}(a) shows how $\langle m\rangle$ approaches 0 as $g$ is increased for different system sizes. Fig. \ref{fig:eth}(b) shows that $\langle m\rangle$ scales with $g$ similar to $\langle\bar{r}\rangle$. The exponent here is $\sim 0.35$, also close to the estimated analytical value.

\begin{figure}[htb]
\vspace{-10pt}
\includegraphics[width=\linewidth]{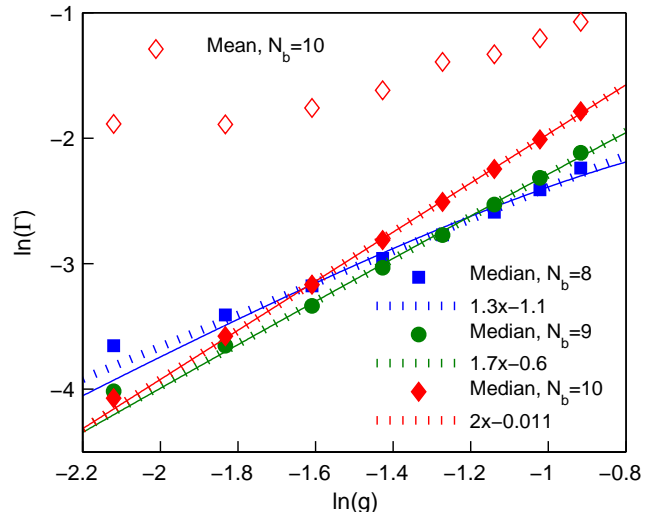}
\vspace{-5mm}
\caption{Broadening of a spectral line as a function of $g$ for a system of $p$-bits with $N=4$ and $N_b=8,9,10$ averaged over more than 38000 eigenstates obtained from several disorder configurations at $w=10$. $\ln(g_c)=-.345N_b<-2.76$ for the sizes shown here. The mean and the median of the probability distribution of the linewidth $\Gamma$ are extracted from the data as discussed in the appendix. The dotted lines are linear fits to the data. The solid lines are fits to the theoretical prediction. 
\label{fig:linewidth} }
\vspace{-10pt}
\end{figure}

We now turn to an analysis of the spectral functions of local operators. Henceforth we are working with the physical degrees of freedom, Eq. (1) and (6). We examine the spectral function from an exact eigenstate
\begin{equation}
A_{i,\alpha}(E)=\Sigma_m|<\psi_{m}|\sigma^x_i|\psi_{\alpha}>|^2\delta_{E_{\psi_{m}}-E_{\psi_{\alpha}},E}
\end{equation}
where $|\psi_m\rangle$ is the $m^{th}$ eigenstate of the combined system and bath. We note that since we are working with a finite size system with a discrete spectrum, the spectral function will always consist of a set of delta functions. At $g=0$, the delta functions should have minimum spacing $2^{-N}$, equal to the many body level spacing in the system. At non-zero $g$, each `parent' delta function will split into exponentially many descendants, with a typical spacing $2^{-N_b}$. A fine binning in energy with bin size greater than $2^{-N_b}$ will then yield a smooth spectral function, with the `parent' delta functions of the system having been `broadened' by coupling to the bath. To investigate this broadening, it is convenient to take $N_b \gg N$. We therefore take $N = 4$ and $N_b = 8,9,10$, and investigate how the `line broadening' evolves with $g$ for $g>g_c$. Details of the procedure are outlined in the supplementary material, and the results are illustrated in Fig. \ref{fig:linewidth} for $w=10$. The mean and median linewidth at a particular value of $g$ are significantly different. This is a result of the long tails in the distribution of the linewidth (see supplement). Fig. \ref{fig:linewidth} shows that at the larger values of $g$ we study, a log-log plot of the median vs $g$ appears to fit well to a straight (dashed) line. For the system sizes that we are able to access, the straight line fit suggests $\Gamma_{median} \sim g^{\gamma}$, where $\gamma $ increases as the size of the bath increases, reaching $2$ for $N_b = 10$.  We note that while a simple application of the golden rule predicts $\gamma = 2$, a more careful analysis \cite{rahul} suggests that the true scaling should be $\Gamma_{typical} \sim g^2 \log(1/g^2)$. The solid lines in Fig. \ref{fig:linewidth} are a fit to this theoretical prediction, and are consistent with the data, except at smallest $g$. The discrepancy at smallest $g$ and the difference between median and mean are worthwhile topics for future work. 

Finally, we analyze the behavior of the spectral function averaged over all sites and eigenstates of the system, for $N=N_b=7$. We note that the Hamiltonian (1) has a delocalization-localization phase transition at $w=7$. Fig. \ref{fig:pbits}(a) shows $A(E)$ on the delocalized side of the transition for a small value of $g$. $A(E)$ is smooth everywhere. (The graininess is a result of the small system size.)  Fig. \ref{fig:pbits}(b) is on the localized side of the transition, with the system almost decoupled from the bath. Here, $A(E)$ consists of clusters of narrow spectral lines, with a hierarchy of energy gaps, just as was shown to be the case for $l$-bit spectral functions in \cite{rahul}. $A(E)$ vanishes at $E=0$. Thus, local spectral functions can distinguish between extended and localized phases. In Fig. \ref{fig:pbits}(c-e) we examine how the $p$-bit spectral functions evolve  as $g$ increases. We see that the line broadening increases and different lines start to overlap with each other, washing out the weaker spectral features, but larger gaps remain. The zero-frequency gap also fills in with increasing $g$. The spectral functions retain signatures of localization even for $g = 0.2$ when the eigenstates of the combined system and bath are effectively thermal, and get washed out when $g$ becomes comparable to the characteristic energy scales in the system (i.e. $g\sim 1$).

In conclusion, we have investigated the signatures of localization in a disordered system weakly coupled to a heat bath using exact diagonalization. The wave functions are found to exhibit a crossover to thermalization as a function of coupling to the bath. The crossover coupling is proportional to the many body level spacing in the bath, and vanishes exponentially fast in the limit of a large bath size. In contrast, the spectral functions of local operators are found to show more robust signatures of proximity to a localized phase. While the spectral functions are smooth and continuous in the delocalized phase (after coarse graining on the scale of the many body level spacing), the spectral functions in the localized phase consist of narrow spectral lines, and contain a hierarchy of gaps, as well as a gap at zero frequency that persists even after spatial averaging. Increasing the coupling to the bath increases the line broadening (in a manner that we calculate) and washes out these features. However, signatures of localization survive in the spectral functions even at couplings to the bath where the exact eigenstates are effectively thermal (Fig. 1).  

\textit{Acknowledgments:} RN would like to thank Sarang Gopalakrishnan and David Huse for a collaboration on related ideas. This work was supported by DOE grant DE-SC0002140. RNB. acknowledges the hospitality of the Institute for Advanced Study, Princeton while this work was being done. RN was supported by a PCTS fellowship. SJ was supported by the Porter Ogden Jacobus Fellowship of Princeton University.

\section{Appendix}
\begin{figure}[htb]
\includegraphics[width=0.88\linewidth]{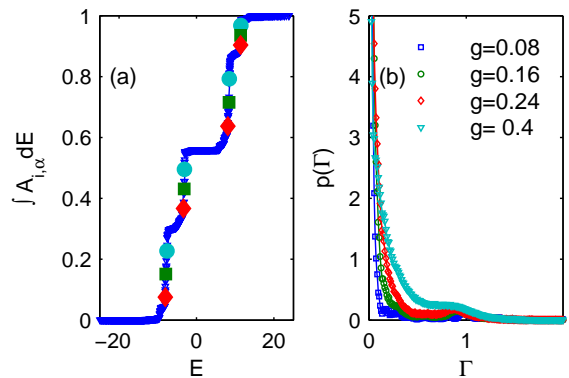}
\vspace{-1mm}
\caption{\label{steps} (a) The procedure for determining the linewidth. The blue curve is an integrated spectral function. The green squares divide each step into half, the red diamonds mark $25\%$ and the light blue circles mark $75 \%$ of each step.  (b) Probability distribution of the linewidth $\Gamma$ for different values of coupling to the bath $g$ for a system with $N=4$ and $N_{\text{b}}=9$ averaged over 10 disorder configurations. Lines are a guide to the eye.}
\label{fig:linewidth2}
\vspace{-0pt}
\end{figure}

In this appendix, we explain how the line width was extracted from the numerical data. We begin by determining the spectral function, defined by 
\begin{equation}
A_{i,\alpha}(E)=\sum_m<\psi_{m}|\sigma^x_i|\psi_{\alpha}>\delta_{E_{\psi_{m}}-E_{\psi_{\alpha}},E}.
\end{equation}
This consists of a set of delta functions. We then define the integrated spectral function $K(E) = \int_{-\infty}^E A(E') dE'$. This consists of a set of step functions (see { Fig. \ref{steps}(a)}). For each step, we identify the energy values corresponding to $25\%$ of the step, $50\%$ of the step, and $75\%$ of the step. The energy spacing between the $25\%$ and $75\%$ points is taken to be the linewidth of this spectral line. We track how this line width scales with $g$. We note that there is in general a wide distribution of line widths for any $g$ (Fig. \ref{steps}(b)). As a result, the mean and the median linewidth scale very differently (see Fig.5 of the main text). An understanding of the difference between the scaling of the mean and typical line width is an important challenge for future work. 

\end{document}